\documentclass{article}%
\usepackage{amssymb}
\usepackage{frascatiphys,here,graphicx,subfigure}
\usepackage{amsmath}
\usepackage{amsfonts}
\usepackage{graphicx}%
\begin{document}

\title{Two-photon decay of light scalars: a comparison of tetraquark and quarkonium assignments}
\author{Francesco Giacosa\\\emph{Institut f\"ur Theoretische Physik} \emph{Universit\"at Frankfurt Johann
Wolfgang}\\\emph{Goethe - Universit\"at Max von Laue--Str. 1 60438 Frankfurt, Germany} }
\maketitle

\begin{abstract}
Two-photon decays of light scalar mesons are discussed within the quarkonium
and tetraquark asignements: in both cases the decay rate of the sigma
resonances turns out to be smaller than 1 keV.

\end{abstract}

\baselineskip=11.6pt

\baselineskip=14pt

\section{Introduction}

The two-photon decay of light mesons represents an important source of
information\cite{amslerrev}. In particular, the $\gamma\gamma$ decay of light
scalar mesons has been considered as a possible tool to deduce their nature.
According to the interpretation of light scalars (quarkonia, tetraquarks or
molecules) different $\gamma\gamma$-rates are expected\cite{pen,tq}. In this
proceeding we first study (Section 2) the quarkonium assignments for the light
scalar states by studying $SU(3)$-relations for the two-photon decays. While
the quarkonium assignment is disfavored when looking at mass
pattern\cite{jaffeorig}, strong decays\cite{tq,maiani} and large-$N_{c}$
studies\cite{pelaez} (see the discussion in the recent
proceeding\cite{ericeproc}), here we adopt a neutral point of view. After a
test-study with the well-known pseudoscalar and tensor mesons\cite{tensor} we
turn to the $\gamma\gamma$ decay of light scalars as quarkonia, finding that
$f_{0}(600)$ has a decay rate well below 1 keV. This is in agreement with the
microscopic evaluation in the recent work\cite{sgg}, where the two-photon
decay of a low-lying quarkonium state with $\overline{n}n=\sqrt{1/2}%
(\overline{u}u+\overline{d}d)$: contrary to usual results quoted in the
literature it is shown that the corresponding two-photon decay rate is well
below $1$ keV for a mass $M_{\overline{n}n}<0.8$ GeV. In Section 3 we turn to
the two-photon transition within the tetraquark assignment and in Section 4 we
briefly summarize our results.

\section{Quarkonia into $\gamma\gamma$}

\subsection{Quarkonia into $\gamma\gamma$: symmetry relations}

We first consider the two-photon decay of scalar quarkonia states. However,
the formula we introduce are also valid, with simple changes, for the
pseudoscalar, tensor and axial vector nonets as we will discuss later on. The
charge neutral scalar quarkonia states $N$, $S$ and $a_{0}^{0}$ are introduced
as $N\equiv\overline{n}n=\sqrt{1/2}(\overline{u}u+\overline{d}d),$
$S\equiv\overline{s}s$ and $a_{0}^{0}\equiv\sqrt{1/2}(\overline{u}%
u-\overline{d}d).$ The $3\times3$ diagonal matrix $\mathcal{S}^{[\overline
{q}q]}\equiv diag\{\overline{u}u,\overline{d}d,\overline{s}s\}=diag\{N/\sqrt
{2}+a_{0}^{0}/\sqrt{2},N/\sqrt{2}-a_{0}^{0}/\sqrt{2},S\}$ plays a central
role. In flavor (and large-$N_{c}$) limit the two-photon decay of these states
is described by the effective interaction Lagrangian%
\begin{equation}
\mathcal{L}_{\gamma\gamma}=c_{\gamma\gamma}Tr\left[  Q^{2}\mathcal{S}%
^{[\overline{q}q]}\right]  F_{\mu\nu}^{2} \label{leffgg}%
\end{equation}
where $Q=$ $diag\{2/3,-1/3,-1/3\}$ is the quark charge matrix, $F_{\mu\nu
}=\partial_{\mu}A_{\nu}-\partial_{\nu}A_{\mu}$ the field tensor of the
electromagnetic field $A_{\mu}$ and $c_{S\gamma\gamma}$ a coupling constant.
As a result the decay rates of $N,$ $S,$ and $a_{0}^{0}$ are given by%
\begin{equation}
\Gamma_{N\gamma\gamma}=\frac{c_{\gamma\gamma}^{2}}{4\pi}M_{N}^{3}\left[
\frac{5}{9\sqrt{2}}\right]  ^{2},\,\Gamma_{S\gamma\gamma}=\frac{c_{\gamma
\gamma}^{2}}{4\pi}M_{S}^{3}\left[  \frac{1}{9}\right]  ^{2},\,\Gamma
_{a_{0}^{0}\gamma\gamma}=\frac{c_{\gamma\gamma}^{2}}{4\pi}M_{a_{0}^{0}}%
^{3}\left[  \frac{3}{9\sqrt{2}}\right]  ^{2}. \label{rels1}%
\end{equation}
The physical states, denoted as $f_{0}(600)$ and $f_{0}(980)$ in the
low-scalar case, are in general a mixing of $N$ and $S$: $f_{0}(600)=\cos
\varphi_{S}N+\sin\varphi_{S}S$ and orthogonal combination for $f_{0}(980)$.
The two-photon decay rates of $f_{0}(600)$ and $f_{0}(980)$ are given by:%
\begin{align*}
\Gamma_{f_{0}(600)\gamma\gamma}  &  =\frac{c_{\gamma\gamma}^{2}}{4\pi}%
M_{f_{0}(600)}^{3}\left[  \frac{5}{9\sqrt{2}}\cos\varphi_{S}+\frac{1}{9}%
\sin\varphi_{S}\right]  ^{2},\,\\
\Gamma_{f_{0}(980)\gamma\gamma}  &  =\frac{c_{\gamma\gamma}^{2}}{4\pi}%
M_{f_{0}(980)}^{3}\left[  -\frac{5}{9\sqrt{2}}\sin\varphi_{S}+\frac{1}{9}%
\cos\varphi_{S}\right]  ^{2}.
\end{align*}
Before studying the scalar case we test these simple expressions on other
nonets. In particular, we will be interested to the ratios $\frac
{\Gamma_{f_{0}(600)\gamma\gamma}}{\Gamma_{a_{0}^{0}\gamma\gamma}}$ and
$\frac{\Gamma_{f_{0}(980)\gamma\gamma}}{\Gamma_{a_{0}^{0}\gamma\gamma}}$ for
which the dependence on the unknown parameter $c_{\gamma\gamma}$ cancels out
under the hypothesis that an eventual momentum dependence is weak, see the
following discussions.

\emph{Pseudoscalar nonet-} Here we have $\pi^{0}=\sqrt{1/2}(\overline
{u}u-\overline{d}d)$ and the isoscalar states $\eta$ and $\eta^{\prime},$
expressed in terms of bare states as%
\begin{equation}
\left(
\begin{array}
[c]{c}%
\eta\\
\eta^{\prime}%
\end{array}
\right)  =\left(
\begin{array}
[c]{cc}%
\cos\varphi_{P} & \sin\varphi_{P}\\
-\sin\varphi_{P} & \cos\varphi_{P}%
\end{array}
\right)  \left(
\begin{array}
[c]{c}%
N\\
S
\end{array}
\right)  =\left(
\begin{array}
[c]{cc}%
\cos\theta_{P} & -\sin\theta_{P}\\
\sin\theta_{P} & \cos\theta_{P}%
\end{array}
\right)  \left(
\begin{array}
[c]{c}%
P^{8}\\
P^{0}%
\end{array}
\right)  \label{etaetap}%
\end{equation}
where $P^{8}=\sqrt{1/6}(\overline{u}u+\overline{d}d-2\overline{s}s)$ and
$P^{0}=\sqrt{1/3}(\overline{u}u+\overline{d}d+\overline{s}s).$ In the
literature the angle $\theta_{P}$ is in general discussed. The corresponding
Lagrangian is similar to (\ref{leffgg}): $\mathcal{L}_{P\gamma\gamma
}=c_{P\gamma\gamma}Tr\left[  Q^{2}\mathcal{P}^{[\overline{q}q]}\right]
F_{\mu\nu}\widetilde{F}^{\mu\nu},$ where $\widetilde{F}^{\mu\nu}%
=\varepsilon^{\mu\nu\rho\sigma}F_{\rho\sigma}$: eq. (\ref{rels1}) retain the
same form. Out of (\ref{etaetap}) one finds the relation $\theta_{P}%
=-(\arcsin[\sqrt{2/3}]+\varphi_{P})=-(54.736^{\circ}+\varphi_{P})$.
In\cite{pdg} the values $\Gamma_{\pi^{0}\gamma\gamma}=7.74\pm0.55$ eV,
$\Gamma_{\eta\gamma\gamma}=0.510\pm0.026$ keV and $\Gamma_{\eta^{\prime}%
\gamma\gamma}=4.30\pm0.15$ keV. The corresponding experimental ratios read
$\frac{\Gamma_{\eta\gamma\gamma}}{\Gamma_{\pi^{0}\gamma\gamma}}=65.9\pm8.1,$
$\frac{\Gamma_{\eta^{\prime}\gamma\gamma}}{\Gamma_{\pi^{0}\gamma\gamma}%
}=556\pm59$. A fit of $\varphi_{P}$ to the latter ratios implies $\varphi
_{P}=-36.0^{\circ},$ and thus $\theta_{P}=-18.7^{0}.$ The corresponding ratios
evaluated at this mixing angle read $\frac{\Gamma_{\eta\gamma\gamma}}%
{\Gamma_{\pi^{0}\gamma\gamma}}=76.6$ and $\frac{\Gamma_{\eta^{\prime}%
\gamma\gamma}}{\Gamma_{\pi^{0}\gamma\gamma}}=661.9$ with $\chi^{2}/2=2.48.$
Taking into account that we are considering the easiest possible scenario,
thus neglecting many possible corrections, these results are very good. In
fact, the experimental results range within 3 order of magnitudes: the
theoretical dependence on the third power of the mass is well verified. An
eventual mass dependence of the effective coupling $c_{P\gamma\gamma}$ has to
be small form $M_{\pi}$ up to 1 GeV, a remarkable fact. The mixing angle
$\theta_{P}=-18.7^{0}$ is in the phenomenological range between $-10^{\circ}$
and $-20^{\circ},$ as found also by more refined studies. This simple exercise
shows that SU(3) flavor relations, together with the $OZI$ rule allowing to
include the flavor singlet in the game, is well upheld and a good starting
point to test the quarkonium assignment in a given nonet.

\emph{Tensor nonet}- As a further test let's consider the tensor nonet
(see\cite{tensor} and refs. therein). The resonances under study are the
isovector $a_{2}(1320)$ with $M_{a_{2}}=1318.3$ MeV and the isoscalars
$f_{2}(1270)$ and $f_{2}^{\prime}(1525)$ with $M_{f_{2}}=1275.1$ and
$M_{f_{2}}=1525$ (we omit tiny errors, see\cite{pdg}). As before, the mixing
angle $\varphi_{T}$ is introduced as $f_{2}(1270)=\cos\varphi_{T}N+\cos
\varphi_{T}S$ and orthogonal combination for $f_{2}^{\prime}(1525)$. The
interaction Lagrangian is $\mathcal{L}_{T\gamma\gamma}=c_{T\gamma\gamma
}Tr\left[  Q^{2}\mathcal{T}_{\mu\nu}^{[\overline{q}q]}\right]  \Theta^{\mu\nu
}$ where $\Theta^{\mu\nu}$ is the energy-momentum tensor of the
electromagnetic fields. The experimental values $\Gamma_{a_{2}\gamma\gamma
}=1.00\pm0.06$ keV, $\Gamma_{f_{2}\gamma\gamma}=2.60\pm0.24$ keV and
$\Gamma_{f_{2}^{\prime}\gamma\gamma}=(8.1\pm0.9)10^{-2}$ keV lead to the
ratios $\frac{\Gamma_{f_{2}\gamma\gamma}}{\Gamma_{a_{2}\gamma\gamma}}%
=2.6\pm0.4$ and $\frac{\Gamma_{f_{2}^{\prime}\gamma\gamma}}{\Gamma
_{a_{2}\gamma\gamma}}=\left(  8.1\pm1.4\right)  10^{-2}$. A fit of the angle
$\varphi_{T}$ to these values leads to $\varphi_{T}=8.19^{\circ}$ and a very
small $\chi^{2}/2=0.015,$ thus the experimental values are reproduced almost
exactly. The value of $\varphi_{T}=8.19^{\circ}$ is in good agreement with the
study of strong decays\cite{tensor}. Again, the $\gamma\gamma$-ratios can be
well described by simple symmetry relations. Indeed, the agreement is even
better than in the pseudoscalar case: this is expected because the masses vary
in a smaller energy region. In the end, we remind that a quarkonium
interpretation works well for vector mesons: here the dominant e.m. transition
is into one single photon (i.e. mixing) which is at the basis of the
successful phenomenology of the vector meson dominance hypothesis.

\emph{Scalar nonet below 1 GeV-} Let us now turn back to the scalar states
below 1 GeV within a quarkonium assignment. We identify the neutral isovector
$a_{0}^{0}$ with $a_{0}^{0}(980)$ and, as described above, the isoscalars with
$f_{0}(600)$ and $f_{0}(980).$ The experimental results for the decay width of
$a_{0}$ and $f_{0}(980)$ are given by\cite{pdg}: $\Gamma_{f_{0}(980)\gamma
\gamma}=0.39_{-0.13}^{+0.10}$ KeV, $\Gamma_{a_{0}^{0}\gamma\gamma}%
=0.30\pm0.10$ KeV. Thus, the experimental ratio reads then $\frac
{\Gamma_{f_{0}(980)\gamma\gamma}}{\Gamma_{a_{0}^{0}\gamma\gamma}}=1.30\pm0.8$
where an average error of $0.115$ KeV for $\Gamma_{f_{0}\gamma\gamma}$ has
been used. As noticeable, the error for this ratio is large. Unfortunately,
the experimental situation concerning $f_{0}(600)\rightarrow2\gamma$ is even
worse; no average or fit is presented in\cite{pdg}, however two experiments
listed in\cite{pdg} find large $\gamma\gamma$ decay widths: $3.8\pm1.5$ keV
and $5.4\pm2.3$ keV, respectively. In a footnote it is then state that these
values could be assigned to $f_{0}(1370)$ (actually, in a older version of
PDG\cite{pdg2000} these values were assigned to the resonance $f_{0}(1370)$).
It is not clear if the $\gamma\gamma$ signal comes from the high mass tail of
the broad $\sigma$ state or from $f_{0}(1370)$ (or even from both). We
determine the mixing angle $\varphi_{S}$ by using the experimental result
$\frac{\Gamma_{f_{0}(980)\gamma\gamma}}{\Gamma_{a_{0}^{0}\gamma\gamma}%
}=1.30\pm0.8$. Due to the large error we report the possible ranges for
$\varphi_{S}$ compatible with it: $-105.9^{\circ}\leq\varphi_{S}%
\leq-47.4^{\circ}$ (central value $\varphi_{S}=-56.9^{\circ}$) and
$15.9^{\circ}\leq\varphi_{S}\leq74.2^{\circ}$ (central value $\varphi
_{S}=25.3^{\circ}$). Indeed, as we saw previously the angle $\varphi_{P}$ in
the pseudoscalar sector is negative: the components of $N$ and $S$ are in
phase for $\eta^{\prime}$ and out of phase for $\eta$ ($\varphi_{P}%
=-36.0^{\circ}$). Within the NJL model the mixing strength is generated by the
't Hooft term, and turns out to have opposite sign with respect to the
pseudoscalar sector: this would favour a positive value of $\varphi_{S}.$
Furthermore, studies without the strange meson can reproduce the $f_{0}(600)$
resonance: this favors small mixing angles. Then, a value $\varphi_{S}%
\sim25^{\circ}$ is favoured. The corresponding two-photon decay rate is
$\frac{\Gamma_{f_{0}(980)\gamma\gamma}}{\Gamma_{a_{0}^{0}\gamma\gamma}}%
\sim0.4$ (which in turn means $\Gamma_{f_{0}(980)\gamma\gamma}\sim0.12$ KeV,
considerably smaller than the above mentioned (but not established)
experimental result). Surely, when including finite width effects the decay
rate $\Gamma_{f_{0}(600)\gamma\gamma}$ increases: in fact, the kinematical
factor $M_{f_{0}(600)}^{3}$ makes the right-tail of the broad distribution of
$f_{0}(600)$ important\cite{tq,pagliara}. However, the rate of increase is not
dramatic: it can at most double the above quoted results but not reach values
of about 2-5 keV. This result is contrary to the usual belief that a
quarkonium decay rate should be well above 1 keV\cite{pen,scadron}: indeed, as
we discuss in the next subsection a careful microscopic calculation of the
two-photon decay rate shows that results below 1 keV are expected\cite{sgg}.

For the discussion of scalar states above 1 GeV we refer to\cite{close}, where
the inclusion of a glueball state mixing with quarkonia also influences the
$\gamma\gamma$ decay rate: in fact, a glueball is expected to have a small
$\gamma\gamma$-transition amplitude, thus if a resonance will have a
consistent glueball component the $\gamma\gamma$ decay rate is small. No
$\gamma\gamma$-signal is found for $f_{0}(1500)$ pointing to a large gluonic
amount in its wave function.

\subsection{Quarkonium into $\gamma\gamma$: a microscopic evaluation}

In this subsection we refer to\cite{sgg}, where the $\gamma\gamma$-decay rate
has been carried out within a local and nonlocal microscopic model. Here we
discuss only the latter. The relevant nonlocal interaction Lagrangian,
involving the mesonic quarkonium field $\sigma(x)\,$ and the quark fields
$q^{t}=(u,d),$ reads\cite{sgg,faessler}
\begin{equation}
\mathcal{L}_{\mathrm{int}}(x)\,=\,\frac{g_{\sigma}}{\sqrt{2}}\sigma(x)\,\int
d^{4}y\,\Phi(y^{2})\,\bar{q}(x+y/2)q(x-y/2)\,, \label{lnonloc}%
\end{equation}
where the delocalization takes account of the extended nature of the
quarkonium state by the covariant vertex function $\Phi(y^{2})$. The
(Euclidean) Fourier transform of this vertex function is taken as
$\widetilde{\Phi}(k_{E}^{2})=\exp(-k_{E}^{2}/\Lambda^{2}),$ also assuring
UV-convergence of the model. The cutoff parameter $\Lambda$ is vaired between
$1$ and $2$ GeV, corresponding to an extension of the $\sigma$ of about
$l\sim1/\Lambda\sim0.5$ fm. The coupling $g_{\sigma}$ is determined by the
so-called compositeness condition\cite{sgg,faessler}. In the calculation the
quark mass varies between 0.3 and 0.45 GeV.

The two-photon decay occurs via a triangle-diagram of quarks. Notice that due
to the presence of the vertex function $\Phi(y^{2})$ inclusion of the
electromagnetic interaction is achieved by gauging the nonlocal interaction
Lagrangian (\ref{lnonloc}): in addition to the usual photon-quark coupling
obtained by minimal substitution new vertices arise, where the photon couples
directly to the $\sigma\gamma\gamma$ interaction vertex, see\cite{faessler}
for details. Their contribution, important on a conceptual level to assure
gauge invariance, is numerically suppressed. In Table~1 we summarize our
results for $M_{\sigma}=0.6$ GeV varying $m_{q}$ both for $\Lambda=1$ GeV and,
in parenthesis, for $\Lambda=2$ GeV.

\begin{center}
\textbf{Table 1: }$\Gamma_{\sigma\gamma\gamma}$ for $m_{q}=0.31-0.45$ GeV,
$\Lambda=1(2)$ GeV at $M_{\sigma}=0.6$ GeV.%

\begin{tabular}
[c]{|l|l|l|l|l|}\hline
$m_{q}$ (GeV) & \ 0.31 & \ 0.35 & \ 0.40 & \ 0.45\\\hline%
\begin{tabular}
[c]{l}%
$\Gamma_{\sigma\gamma\gamma}$ (keV)\\
at $M_{\sigma}=0.6$ GeV
\end{tabular}
&
\begin{tabular}
[c]{l}%
0.529\\
(0.512)
\end{tabular}
&
\begin{tabular}
[c]{l}%
0.458\\
(0.415)
\end{tabular}
&
\begin{tabular}
[c]{l}%
0.361\\
(0.327)
\end{tabular}
&
\begin{tabular}
[c]{l}%
0.294\\
(0.267)
\end{tabular}
\\\hline
\end{tabular}

\end{center}

The decay widths decrease slowly for increasing quark mass while the
dependence on the cutoff $\Lambda$ is very weak. The numerical analysis shows
that $\Gamma_{\sigma\gamma\gamma}<1$ keV for $M_{\sigma}<0.7$-$0.8$ GeV$\,$.
This result is indeed in agreement with that of the previous subsection: a
light quarkonium state has a $\gamma\gamma$ decay rate smaller than 1 keV.
However, this doesn't prove that the resonance $f_{0}(600)$ is a
quark-antiquark state. It rather tells us that, being the $\gamma\gamma$ decay
width of a quarkonium smaller than what usually believed, care is needed when
using $\gamma\gamma$-rates to discuss the nature of light scalars. We also
refer to\cite{giacosagl} for the evaluation of these diagrams for quarkonia
states above 1 GeV.

\section{Tetraquarks into $\gamma\gamma$}

We consider now the $\gamma\gamma$-transition of tetraquark states\cite{tq},
whose effective Lagrangian reads%
\begin{equation}
\mathcal{L}_{em}=c_{1}^{\gamma\gamma}\mathcal{S}_{ij}^{[4q]}\left\langle
A^{i}QA^{j}Q\right\rangle F_{\mu\nu}^{2}-c_{2}^{\gamma\gamma}\mathcal{S}%
_{ij}^{[4q]}\left\langle A^{i}A^{j}Q^{2}\right\rangle F_{\mu\nu}^{2},
\label{laggg}%
\end{equation}
where $\left(  A^{i}\right)  _{jk}=\varepsilon_{ijk}$ and $\mathcal{S}%
^{[4q]}=diag\{\sqrt{\frac{1}{2}}(f_{B}^{[4q]}-a_{0}^{0[4q]}),\sqrt{\frac{1}%
{2}}(f_{B}^{[4q]}+a_{0}^{0[4q]}),\sigma_{B}^{[4q]}\}$. Within the tetraquark
assignment the isoscalars are $f_{B}^{[4q]}=\frac{\left(  \left[  u,s\right]
\left[  \overline{u},\overline{s}\right]  +\left[  d,s\right]  \left[
\overline{d},\overline{s}\right]  \right)  }{2\sqrt{2}},$ $a_{0}^{0[4q]}%
=\frac{\left(  \left[  u,s\right]  \left[  \overline{u},\overline{s}\right]
-\left[  d,s\right]  \left[  \overline{d},\overline{s}\right]  \right)
}{2\sqrt{2}}$ and $\sigma_{B}^{[4q]}=\frac{1}{2}\left[  u,d\right]  \left[
\overline{u},\overline{d}\right]  $. In eq. (\ref{laggg}) two terms are
present: the one proportional to $c_{1}^{\gamma\gamma}$ represents the
dominant contribution in the large-N$_{c}$ expansion (switch of a quark with
an antiquark), while the second term, proportional to $c_{2}^{\gamma\gamma}$
(annihilation of a quark-antiquark pair), represents the next-to-leading order
correction. As discussed in detail in\cite{tq,ericeproc,tqmix} the latter
mechanism can be relevant because it occurs with only one gluon as
intermediate state. The decay width into two photons reads $\Gamma
_{i\gamma\gamma}=\frac{M_{i}^{3}}{4\pi}g_{i\gamma\gamma}^{2}$ where
$i=a_{0}^{0[4q]},\sigma_{B}^{[4q]},f_{B}^{[4q]}.$ The coupling constants for
$a_{0}^{0}$ and for the bare states $\sigma_{B}$ and $f_{B}$ are deduced from
(\ref{laggg}) and read:%
\begin{equation}
g_{a_{0}^{0[4q]}\gamma\gamma}=\frac{2c_{1}^{\gamma\gamma}+c_{2}^{\gamma\gamma
}}{3\sqrt{2}},\,\,g_{\sigma_{B}^{[4q]}\gamma\gamma}=\frac{4c_{1}^{\gamma
\gamma}+5c_{2}^{\gamma\gamma}}{9},\,\,g_{f_{B}^{[4q]}\gamma\gamma}%
=\frac{2c_{1}^{\gamma\gamma}+7c_{2}^{\gamma\gamma}}{9\sqrt{2}} \label{tqg}%
\end{equation}
The mixed physical states $f_{0}(600)$ and $f_{0}(980)$ are expressed in the
tetraquark framework as $f_{0}(600)=\cos\varphi_{S}\sigma_{B}^{[4q]}%
+\sin\varphi_{S}f_{B}^{[4q]}$ and orthogonal combination for $f_{0}(980)$. Let
us first consider $c_{2}^{\gamma\gamma}=0.$ When determining the mixing angle
$\varphi_{S}\ $by using the experimental ratio $\frac{\Gamma_{f_{0}%
(980)\gamma\gamma}}{\Gamma_{a_{0}^{0}\gamma\gamma}}=1.30\pm0.8$ one obtains
very large values: $\left\vert \varphi_{S}\right\vert \gtrsim70^{\circ}$
(indeed $\Gamma_{f_{0}(980)\gamma\gamma}/\Gamma_{a_{0}^{0}\gamma\gamma}\leq1$
for each $\varphi_{S}$). One of the main advantages of the tetraquark
assignment is the explanation of the mass degeneracy of $a_{0}(980)$ and
$f_{0}(980)$ in the limit $\varphi_{S}=0.$ However, a large mixing angle would
completely spoil the mass degeneracy. We thus consider this possibility
disfavored, see discussion in\cite{tq}. When $c_{2}^{\gamma\gamma}\neq0$ a
determination of the parameters form the $\gamma\gamma$-data is no longer
possible. However, the mixing angle $\varphi_{SS}$ can be fixed from strong
decays\cite{tq}: $\varphi_{S}=-12.8^{\circ}$. Then we find $0.15\leq
c_{2}^{\gamma\gamma}/c_{1}^{\gamma\gamma}\leq1.39$. Notice that even a small
but non vanishing $c_{2}^{\gamma\gamma}$ can improve a lot the phenomenology:
in fact, $c_{2}^{\gamma\gamma}$ strongly enhances the amplitude $g_{f_{B}%
^{[4q]}\gamma\gamma}$, see eq. (\ref{tqg}). For $0.15\leq c_{2}^{\gamma\gamma
}/c_{1}^{\gamma\gamma}\leq1.39$ one has $\frac{\Gamma_{f_{0}(600)\gamma\gamma
}}{\Gamma_{a_{0}^{0}\gamma\gamma}}\leq0.35,$ again pointing to a small
$\gamma\gamma$-rate of $f_{0}(600)$ as in the quarkonium case. More work is
needed but one result is stable: the $\Gamma_{f_{0}(600)\gamma\gamma}$ is well
below 1 keV also in the tetraquark assignment and is indeed of the same order
of magnitude of the quarkonium interpretation. One could indeed go further by
including mixing of tetraquark below 1 GeV and quarkonia above 1 GeV: however,
as found in\cite{tqmix} the latter turns out to be small, thus not changing
much the present results.

\section{Conclusions}

In this work we discussed the two-photon transition of light scalar mesons
within the quarkonium and the tetraquark assignments. In both cases the decay
rate $\Gamma_{f_{0}(600)\gamma\gamma}$ is smaller than 1 keV, as confirmed by
a microscopic calculation\cite{sgg} in the quarkonium assignment. These
results render a possible identification of the nature of scalar states using
two-photon decay widths more difficult.

\end{document}